\documentclass[%
preprint,
 amsmath,amssymb,
]{revtex4-1}
\usepackage{graphicx}
\usepackage{dcolumn}
\usepackage{bm}
\usepackage{ifpdf}
\usepackage{epstopdf}
\usepackage{slashed}
\usepackage{amsfonts}
\usepackage{mathrsfs}
\usepackage{amssymb}
\usepackage{multirow}
\usepackage{CJK}
\usepackage{xcolor}

\begin{document}

\preprint{IFCEN SYSU}

\title{Productions of high energy neutrons by interactions between deuteron beam and thick target}

\author{Wu Sun}
\author{Weiwei Qiu}
\author{Jun Su}
\email{sujun3@mail.sysu.edu.cn}
\affiliation{Sino-French Institute of Nuclear Engineering and Technology, \\
Sun Yat-sen University, Zhuhai 519082, China}

\date{\today}

\begin{abstract}
The cross sections of high energy neutron-induced spallation is useful for studying the transmutation of long-life fission products. However, due to the difficulty of obtaining high-energy neutrons, the experimental data are still scarce.
The present work studies the possibility to produce high energy neutrons by interactions between deuteron beam and thick target.
The Geant4 toolkit is applied to simulate the interaction between the deuteron beam and thick target.
An analytical method is also developed to calculate the neutron yields emitted in the interaction between the deuteron beam and thick target.
The input cross section data is not only taken from the TEDNL-2017 library but also calculated by the isospin-dependent quantum molecular dynamics model.
It is indicated that it is possible to produce high energy neutron by deuteron beam interaction with matter.
If one wants to get high energy neutrons, low-Z matter, thin target, and small emission angle may be considered.

\end{abstract}

\maketitle

\section{\label{int}Introduction}

How to deal with a large quantities of high-level radioactive wastes has received much attention in the nuclear power industry \cite{RN1}.
So far, the geological disposal is a common method, for what the radioactive waste is buried deep in the ground \cite{RN2,RN3}.
However, the geological disposal is not perfectly safe, because the change geological structures may cause the leakage of radioactive waste \cite{RN4}.
The transmutation is another proposed method to deal with nuclear waste \cite{RN5,RN6,RN7,RN8,RN9}.
In the 1990s, the Accelerator-Driven Systems (ADS) is put forward to transmute high-level radioactive waste \cite{RN10}.
Accelerator-Driven Systems consists of a proton accelerator, a spallation target, and a subcritical reaction system \cite{RN11,RN12,RN13,RN14,RN15}.

For the ADS, spallation reaction is one of its core reactions to produce high energy neutron.
It is also proposed to be suitable for the transmutation of long-life fission products (LLFPs).
In recent years, some experiments have shown that the spallation reaction induced by proton or deuteron has a good effect on the transmutation of LLFPs \cite{RN16,RN18,RN17}.
Compared with proton and deuteron, the neutron is not affected by the electromagnetic interaction force during the movement.
Neutron beams can trigger more fission reactions at the same energy condition \cite{RN19}.
However, the experimental neutron-induced spallation cross section data of the LLFPs are very scarce, especially in the high energy region.
For the LLFPs elements like $^{129}$I, $^{135}$Cs, and $^{107}$Pd, data is available only at neutron energy below 1 MeV \cite{RN20}.
When the incident energy is greater than 10 MeV, the experimental data are all blank.
One needs high energy neutrons for experiments on the measurement of cross sections.
However, due to the limitation of the existing experimental conditions, the acquisition of high energy neutrons is still very difficult today \cite{RN21,RN22,RN23}.

To get high energy neutrons, many neutron generators have been designed.
A commonly used neutron generator is the D-D neutron generator, where the fusion of deuterium atoms (D + D) results in the neutron with a kinetic energy of approximately 2.5 MeV \cite{RN31}.
The other type is the D-T neutron generator, where the fusion of a deuterium and a tritium atom (D + T) results in higher neutron, i.e. 14.1 MeV \cite{RN32}.
The neutron generators in the isolating core transformer accelerator at Lawrence Livermore National Laboratory \cite{RN24} and Lawrence Berkeley National Laboratory \cite{RN27} are both the D-T type.
Based on the reaction $^{7}$Li(p,n)$^{7}$Be, one can build neutron generator to produce quasi-monoenergetic neutron.
For example, quasi-monoenergetic neutron beams in the energy range from about 25 to 70 MeV were produced at the accelerator facility of the Universit\'{e} Catholique de Louvain \cite{RN25}, while those 40-90 MeV were produced at TIARA of Japan Atomic Energy Research Institute \cite{RN26}.
The cross section of the reaction $^{7}$Li(p,n)$^{7}$Be is tiny for proton energy higher than 100 MeV.
Thus the neutron energy higher than 100 MeV is commonly generated by another reaction type, i.e. spallation.
For example, the China spallation neutron source can provide neutron at hundreds MeV \cite{RN28, RN29, RN30}.
However, based on proton-induced spallation, the neutron flux decreases with increasing energy.

In our previous work, it has been shown that the neutron production cross section in the deuteron-induced spallation is huge near the incident energy \cite{RN19}.
It may be applied to produce high energy neutrons.
In this paper, we study the productions of high energy neutrons by interactions between deuteron beam and thick target.
The paper is organized as follows. In Sec. \ref{method}, the method is shown.
In Sec. \ref{res}, the results and discussions are presented.
In Sec. \ref{con}, we give summaries.

\section{\label{method} Theoretical framework}

\subsection{ Geant4 }

Geant4 is a Monte Carlo toolkit to simulate the particle transport in matter, which is developed by European Nuclear Research Organization \cite{RN33}.
The source code of the Geant4 toolkit is completely open.
Users can change and expand Geant4 programs according to actual needs.
In this work, the version G4.10.05p.01 is performed on the Ubuntu 18.04 system.
The G4TENDL1.3.2 database and G4PARTICLEXS1.1 database are used in the simulation.
Their data comes mostly from TENDL-2014 library, with some isotopes taken from ENDF/B-VII.1\cite{RN34}.
The nuclear data library TENDL-2014 was released on December 11, 2014 \cite{RN35}.
The ENDF/B-VII.1 library was released on December 22, 2011 by the Cross Section Evaluation Working Group.
It is recommended evaluated nuclear data file for use in nuclear science and technology applications\cite{RN36}.
By combining these two databases with the Monte Carlo method, we can use the Geant4 toolkit to analyze the neutron yield of different targets \cite{RN37}.

\subsection{ Analytical calculation of neutron yield }

In an interaction between a deuteron beam with intensity I and a matter with thickness dx, the number of neutrons emitted at the solid angle d$\Omega$ can be expressed as,

\begin{equation}
\frac{d^{2} n}{d \Omega}=I \frac{d \sigma}{d \Omega} \frac{\rho N_{A}}{M} d x
\label{Y-dx}
\end{equation}
where $\frac{d \sigma}{d \Omega}$ is differential cross section, ${\rho}$ presents the density of the matter, M is the molar mass of the matter, N$_{A}$ represents Avogadro's constant.
For the thick matter, the neutron production is calculated by the integral,

\begin{equation}
\frac{d n}{d \Omega}= \int_{0}^{x} I \frac{d \sigma}{d \Omega} \frac{\rho N_{A}}{M} d x,
\label{Y-dx2}
\end{equation}
where x is the thickness of the matter.

The differential equation for the intensity I is expressed as,

\begin{equation}
\frac{d I}{I}=-\sigma_{t} \frac{\rho N_{A}}{M} d x,
\end{equation}
where $\sigma_t$ is the total cross section of the deuteron, which depends on the energy.
The intensity I(x) after the beam passes through the matter with thickness x can be calculated by the integral,
\begin{equation}
\ln [I(x)] =\ln (I_{i}) -\int_{0}^{x}\sigma_{t} \frac{\rho N_{A}}{M} d x,
\label{Ix}
\end{equation}
where I$_{i}$ is the incident intensity.
One can not solve the integrals Eqs. (\ref{Y-dx2}) and (\ref{Ix}), because the both the differential cross section and total cross section depend on the energy E of the beam, while the energy E will change in the matter.

Using the concept of the specific energy loss S, one can relate the thickness dx to the beam energy dE,

\begin{equation}
S=-\frac{d E}{d x} \Rightarrow d x=-\frac{1}{S} dE.
\end{equation}
Then Eqs. (\ref{Y-dx2}) and (\ref{Ix}) can be changed to

\begin{equation}
\frac{d n}{d \Omega}= -\int_{E_{i}}^{E(x)} I(E) \frac{d \sigma}{d \Omega} \frac{\rho N_{A}}{M} \frac{1}{S} d E,
\label{YE}
\end{equation}

\begin{equation}
\ln [I(E)] =\ln (I_{i}) +\int_{E_{i}}^{E(x)} \frac{\rho N_{A}}{M} \frac{1}{S} d E,
\label{IE}
\end{equation}

where E$_i$ represents the incident energy, and E(x) represents the energy as a function of thickness x.
One need the relation E(x) further to solve the integrals.

The specific energy loss S can be calculated by the Bethe Formula,
\begin{equation}
S = \frac{4\pi z^{2} e^{4} }{m_{0} c^{2} \beta^{2} \left(4\pi\varepsilon_{0}\right)^{2}}  \frac{\rho N_{A}Z }{M}
\left[\ln \left(\frac{2 m_{0}c^{2} \beta^{2}}{P}\right)+\ln \left(\frac{1}{1-\beta^{2}}\right)-\beta^{2}\right],
\end{equation}
where z = 1 is the charge number of the deuteron, Z is the charge number of target nucleus, m$_0$ is the electron mass, c is speed of light, $\beta$ is the speed of the deuteron in unit c, P is the average excitation of the target matter.
${\beta}$ can be expressed as a function of energy
\begin{equation}
\beta^{2}=1-\frac{m_{0}^{2} c^{4}}{\left(E+m_{0} c^{2}\right)^{2}}.
\end{equation}
Using the energy dependence of the specific energy loss, the energy as a function of thickness x is calculated by the inverse function E$^{-1}$(x),

\begin{equation}
E^{-1}(x) = -\int_{E_{i}}^{E}\frac{d E}{S(E)}.
\label{xE}
\end{equation}
The physical meaning of the inverse function E$^{-1}$(x) is as following.
The deuteron with energy E$_{i}$ passes through the matter.
With the given outgoing energy E, one can calculate the thickness E$^{-1}$(x) of the matter by Eq. (\ref{xE}).

In fact, Eq. (\ref{YE}) represents neutron production by the primary deuteron-induced reaction.
In order to calculate neutron yield, one need to consider the exponential damping and slowing down of neutrons, as well as the neutron production of the secondly nuclear reactions.
Here, we only consider the exponential damping.

\begin{equation}
\frac{d Y}{d \Omega}= -\int_{E_{i}}^{E(x)} exp\left[ -\frac{x_{t}-E^{-1}(x)}{cos\theta} \sigma_{nt} \frac{\rho N_{A}}{M} \right] I \frac{d \sigma}{d \Omega} \frac{\rho N_{A}}{M} \frac{1}{S} d E,
\label{YE2}
\end{equation}
where x$_{t}$ is the total thickness of the matter, $\sigma_{nt}$ is the total cross section of the neutron, $\theta$ is the emission angle.

With the given differential cross section, deuteron total cross section, and neutron total cross section, one can calculate the neutron yields analytically using Eqs. (\ref{IE}), (\ref{xE}), and (\ref{YE2}).

\subsection{ Isospin dependent quantum molecular dynamics model}

The cross sections are available from the databases such as ENDF/B-VI.8-HE, JENDL/HE-2007 and TENDL-2017.
We also calculate the cross sections by the isospin dependent quantum molecular dynamics (IQMD) model.
It has been shown that the IQMD model is reliable to calculate the cross sections of the spallation \cite{RN17, RN19}.

In the IQMD model, the wave function in the form of a Gaussian wave packet can be used to describe the nucleon in a N-body system. For the N-body system, it can be described by the direct product of these single-nucleon wave functions. By applying the Wigner transform of the quantum wave function, the phase-space density function of the N-body system is calculated by
\begin{equation}f(\mathbf{r}, \mathbf{p}, t)=\sum_{i=1}^{N} \frac{1}{(\pi \hbar)^{3}} e^{-\frac{\left[\mathbf{r}-\mathbf{r}_{i}(t)\right]^{2}}{2 L}} e^{-\frac{\left[\mathbf{p}-\mathbf{p}_{i}(t)\right]^{2} 2 L}{\hbar^{2}}},\end{equation}
where \emph{r$_i$} represents respectively the average position and \emph{p$_i$} is the mean momentum of the i th nucleon, the parameter L is related to the square of the width of the Gaussian wave packet for nucleon. In the N-body system, Hamiltonian consists of three terms,
\begin{equation}H=T+U_{\mathrm{Coul}}+\int V[\rho(\boldsymbol{r})] \mathrm{d} \boldsymbol{r},\end{equation}
where \emph{U$_{Coul}$}  is the Coulomb potential energy and \emph{T} is the kinetic energy. In the third term, the nuclear potential energy density of the asymmetric nuclear matter with density $\rho$ and asymmetry $\delta$ is
\begin{equation}V(\rho, \delta)=\frac{\alpha}{2} \frac{\rho^{2}}{\rho_{0}}+\frac{\beta}{\gamma+1} \frac{\rho^{\gamma+1}}{\rho_{0}^{\gamma}}+\frac{C_{s p}}{2}\left(\frac{\rho}{\rho_{0}}\right)^{\gamma_{i}} \rho \delta^{2},\end{equation}
where $\rho$$_0$ is the normal density. In this paper, $\alpha$ = -356.00 MeV, $\beta$ = -303.00 MeV, $\gamma$=7/6, $C_{sp}$ =38.06 MeV, and $\gamma_{i}$ = 0.75.

The time evolution of the nuclei in the generated mean-field is determined by Hamiltonian equations of motion.
\begin{equation}\dot{\mathbf{r}}_{i}=\nabla_{\mathbf{p}_{i}} H, \quad \dot{\mathbf{p}}_{i}=-\nabla_{\mathbf{r}_{i}} H.\end{equation}

Furthermore, the nucleon-nucleon (NN) collisions are included in the IQMD code, so we can simulate the short-range repulsive residual interaction and describe the random change of the phase space distribution. The differential cross sections of the NN collisions can be calculated by
\begin{equation}\left(\frac{\mathrm{d} \sigma}{\mathrm{d} \Omega}\right)_{i}=\sigma_{i}^{\mathrm{free}} f_{i}^{\mathrm{angl}} f_{i}^{\mathrm{med}},\end{equation}
where $\sigma_{i}^{free}$ refers to the cross section of the NN collision in free space, \emph{f}$_i$$^{angl}$ is the angular distribution, \emph{f}$_i$$^{med}$ is the in-medium factor, and the subscript \emph{i} refers to different channels of the NN collision. For elastic proton-proton scattering, \emph{i = pp}. elastic neutron-proton scattering, \emph{i = np}. elastic neutron-neutron scattering, \emph{i = nn}. inelastic neutron-neutron scattering, \emph{i = in}.

In order to improve the stability of the N-body system, the method of phase space density constraint and the Pauli blocking are used in the IQMD model. For a hypercube of volume \emph{h$^3$} in the phase space revolving around the \emph{i}-th nucleon, by integrating over it at each time step, the space occupation probability \emph{f}$_i$ is calculated as
\begin{equation}\bar{f}_{i}=\sum_{n} \delta_{\tau_{n}, \tau_{i}} \delta_{S_{n}, s_{i}} \int_{h^{3}} \frac{1}{\pi^{3} \hbar^{3}} e^{-\frac{\left(\mathrm{r}-\mathrm{r}_{n}\right)^{2}}{2 L}-\frac{\left(\mathrm{p}-\mathrm{p}_{n}\right)^{2} L}{\hbar^{2}}} d^{3} r d^{3} p,\end{equation}
where $\tau_{i}$ is the isospin degree of freedom and s$_{i}$ is the spin projection quantum number.

\section{\label{res} Results and discussions}
\subsection{Comparison of cross sections obtained by different methods}
\DeclareGraphicsExtensions{.eps,.ps,.jpg,.bmp}
\begin{figure}[htb]
	\centering
	\includegraphics[width=8.6cm]{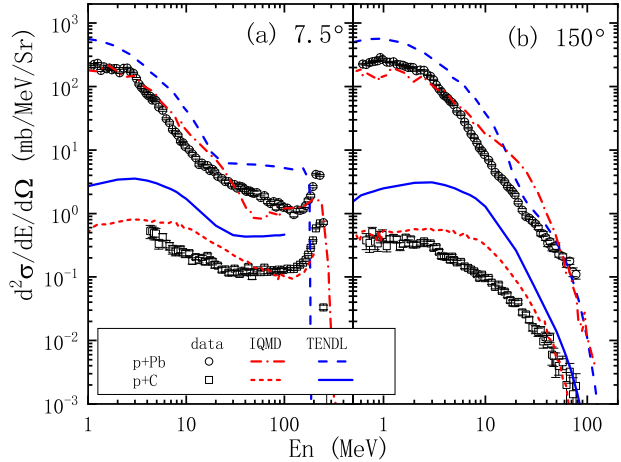}
	\caption{Double differential cross sections of neutrons produced in p + $^{208}$Pb and p + $^{12}$C reactions at 256 MeV. The cross sections are measured and calculated at 7.5$^{\circ}$ and 150$^{\circ}$. The calculations, shown as curves, are obtained by the IQMD model and TENDL database. The data, shown as circles and squares, are taken from Ref \cite{RN38}.}
	\label{fig-1}
\end{figure}

In order to verify the reliability of the IQMD model, it is used to calculate the differential cross sections of neutrons emitted at 7.5$^{\circ}$ and 150$^{\circ}$ in reactions p+$^{208}$Pb and p+$^{12}$C at 256 MeV.
In Fig. \ref{fig-1}, the results are shown and compared to the experimental data and the predictions in the TENDL database.
As the open circles and squares show, the experimental cross sections at 150$^{\circ}$ decrease with increasing neutron energy in the displaying energy region.
For neutrons emitted at small angle (7.5$^{\circ}$), the cross section increases significantly near the incident energy, resulting in a peak at 256 MeV.
The neutron at 256 MeV mainly comes from the central collision.
The neutrons around 1 MeV mostly come from hot nucleus after the collisions.
As shown by the full and dashed lines, the predictions of the TENDL-2017 database fit the trend of data.
However, there are two main deviations.
First, the predicted value overestimates globally the cross section except the case at 30 MeV and 150$^{\circ}$.
The predicted cross section is almost more than twice the experimental data.
For example, for 100 MeV neutrons generated at 7.5$^{\circ}$ for p + $^{12}$C reaction, the experimental data is 0.1 mb/MeV/Sr, while the prediction of the TENDL-2017 database reaches 0.4 mb/MeV/Sr.
Second, there is no obvious peak at 256 MeV.
At present, there are few experimental data of neutron double differential cross section in the high energy region.
The database does not focus on the accuracy for the high energy.
The calculation by the IQMD model reproduces the general trend of the experimental data.
It only overestimates the cross section in the area around 10 MeV.
It is shown a peak at 256 MeV but underestimates the value of peak.
The calculation by the IQMD model is generally better than the predictions of the TENDL-2017 database, especially for the p + $^{12}$C reaction.

\begin{figure}[htb]
	\centering
	\includegraphics[width=8.6cm]{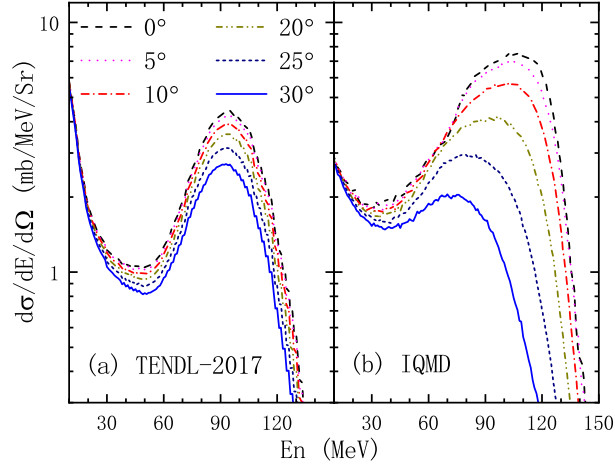}
	\caption{Cross sections of high energy neutrons produced in d + $^{12}$C spallation at 200 MeV at 0$^{\circ}$ to 30$^{\circ}$. (a) The cross sections are taken from TENDL-2017. (b) The cross sections are calculated by IQMD model.}
	\label{fig-2}
\end{figure}

To investigate deeply the difference between the results obtained by IQMD model and the predictions of the TENDL-2017 database, the calculation for d + $^{12}$C reaction at 200 MeV by the IQMD model is compared with the predictions in the TENDL-2017 database at different angles.
In Fig. \ref{fig-2}, the calculation by the IQMD model has the same trend as the predictions of TENDL-2017.
But the calculated values are different.
For example, at 10 MeV, the cross section predicted by TENDL-2017 is 6 mb/MeV/Sr, which is twice that calculated by the IQMD model.
There are peaks near 90 MeV both for TENDL-2017 and IQMD.
However, the angle dependence of the peak is quite different between TENDL-2017 and IQMD.
For TENDL-2017, the peak value decreases from 4.5 to 2.5 mb/MeV/Sr with increasing angle from 0$^{\circ}$ to 30$^{\circ}$.
The peak position is angle independent.
For IQMD, both the peak value and the peak position decrease with increasing angle.
For example, The curve of 30 $^{\circ}$ reaches its peak at a neutron energy of 70 MeV, but for the curve of 0$^{\circ}$, it reaches its peak at around 120 MeV.
This shows that in a smaller angle range, the energy of neutron is higher.
It is proposed to produce high energy neutron at small angle.

\begin{figure}[htb]
	\centering
	\includegraphics[width=8.6cm]{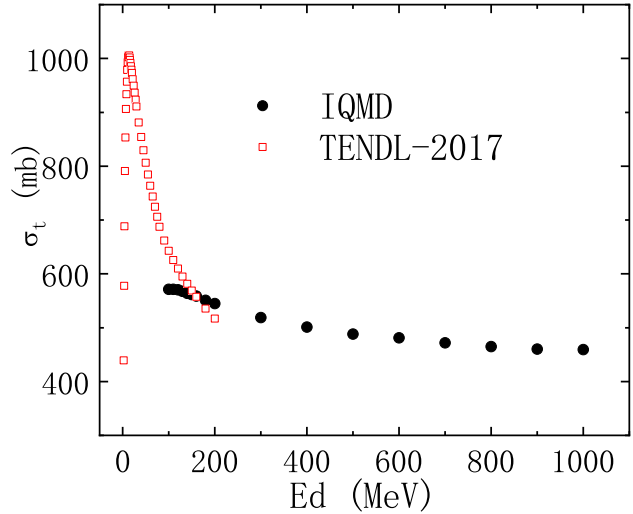}
	\caption{Total cross sections of neutron produced in d + $^{12}$C reaction as a function of projectile energy. The data from TENDL-2017 are shown as open squares. The calculations by the IQMD model are shown as solid circles. }
	\label{fig-3}
\end{figure}

Figure \ref{fig-3} shows the total cross sections of d + $^{12}$C reaction calculated by the IQMD model.
The results are compared to the predictions of TENDL-2017.
The calculations by IQMD model is for projectile energies larger than 100 MeV.
TENDL-2017 gives total cross sections from 0 to 200 MeV.
It can be found in Fig.3 that the total cross section curve rises and then falls.
When the energy of the incident deuteron is about 50 MeV, the total cross section reaches the maximum. Between 100-200 MeV, both the TENDL-2017 and IQMD models can predict the total cross section.
Their predicted total cross sections decrease as the incident deuteron energy increases.
But the predictions of TENDL-2017 decline faster than the results of IQMD model.

\begin{figure}[htb]
	\centering
	\includegraphics[width=8.6cm]{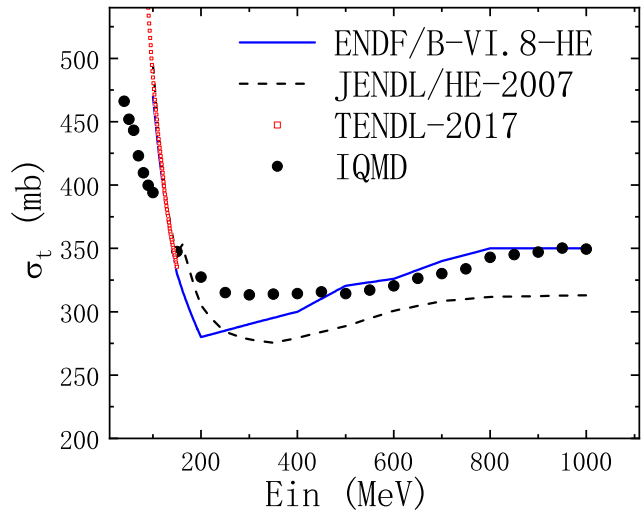}
	\caption{Comparison of the total cross sections predicted by the IQMD model, ENDF/B-VI.8-HE, JENDL/HE-2007 and TENDL-2017 for n + $^{12}$C spallation as a function of projectile energy. The calculations by the IQMD model are shown as a (black) solid circles.  }
	\label{fig-4}
\end{figure}

Figure \ref{fig-4} shows the total cross sections of n + $^{12}$C reaction obtained by four different methods.
By comparing the curves obtained by these four calculation methods, we can see that their trends are similar.
With the rise of neutron energy, the total cross section of the reaction n + $^{12}$C will first decline rapidly and then slowly rise.
The cross section obtained by ENDF/B-VI.8-HE is only for projectile energies more than 100 MeV.
The cross section data from JENDL/HE-2007 is for the projectile energies more than 150 MeV.
The data from TENDL-2017 database agrees well with the result of ENDF/B-VI.8-HE for the projectile energies less than 200 MeV.
The IQMD model gives total cross sections of reaction n + $^{12}$C from 30 to 1000 MeV.
The range of projectile energies is wider than others.
Its results are in good agreement with the predictions of ENDF/B-VI.8-HE for the projectile energies more than 500 MeV.
The common point of these curves is that when the incident neutron energy is low, the total cross section is the largest.
It is indicated that when the energy of the incident particles is low, the probability of a reaction between the particles is greater.

\subsection{Interaction between deuteron beam and matter by Geant4}

\begin{figure}[htb]
	\centering
	\includegraphics[height=8.6cm,width=10cm]{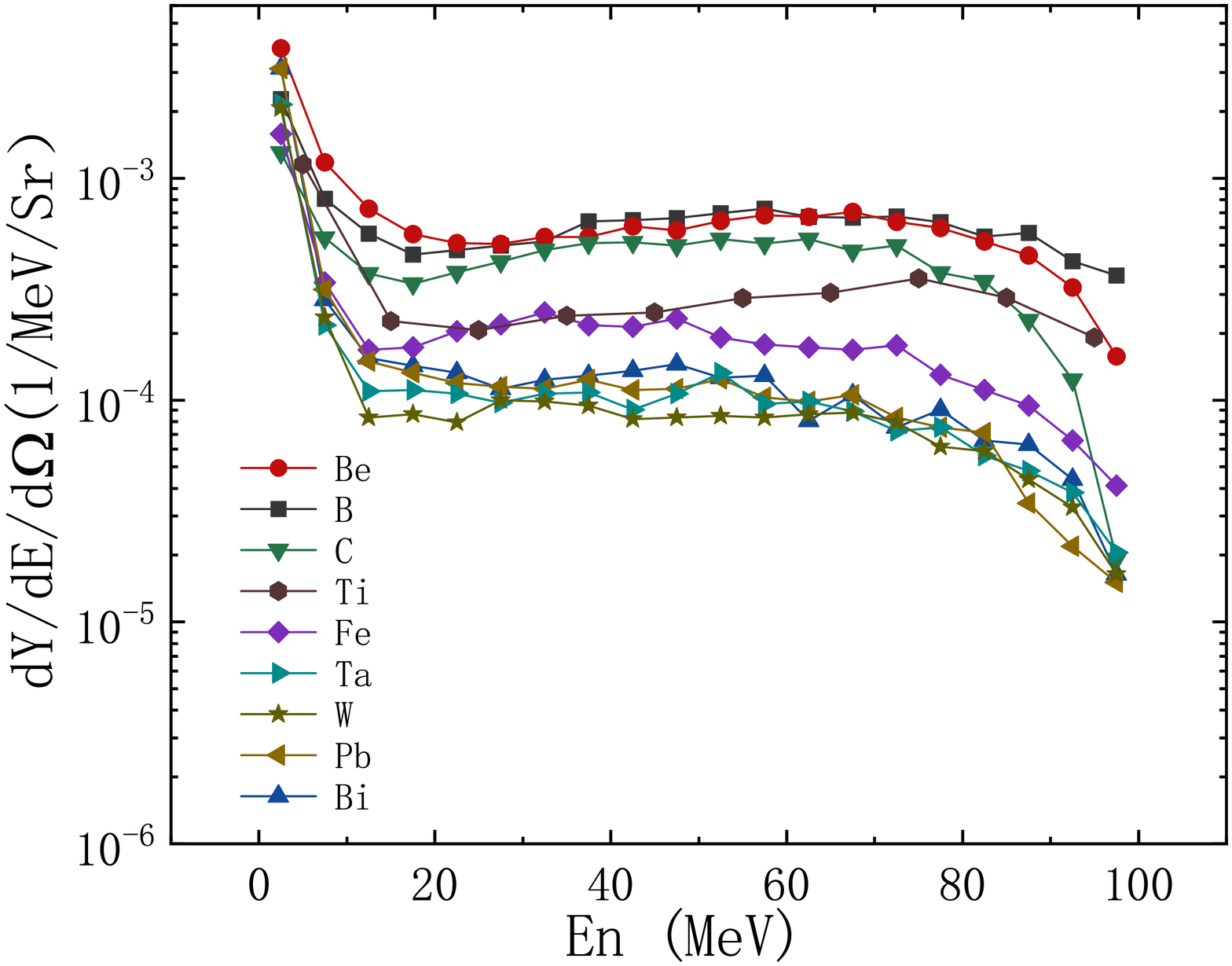}
	\caption{Neutron yields for deuteron beam interaction with different materials, the energy of the deuteron beam is 100MeV. }
	\label{fig-5}
\end{figure}

It is studied the interaction between the deuterium beam and different targets by Geant4.
Figure \ref{fig-5} shows the effect of the target material on neutron energy distribution.
The energy of deuteron beam is 100 MeV.
The target is cylindrical with thickness 2 cm and radius 7 cm.
Nine simple substances are applied for the simulations respectively.
They are Be, B, C, Ti, Fe, Ta, W, Pb, Bi.
The detector is set at the emission angle range from 0 to 16 $^{\circ}$.
The results are shown in Fig. \ref{fig-5}.
The horizontal axis represents the neutron energy.
The vertical axis represents the neutron yield.
In Fig. \ref{fig-5}, it can be found that no matter what kind of material we use, the deuterium interacting causes a lot of low energy neutrons.
Be target produces the most low energy neutrons after the reaction.
The number of high energy neutrons produced by the C target is higher than that for other materials.
It can be also noticed that the neutron yield in the high energy part (70-90 MeV) is related to the atomic number of the material.
When the atomic number Z of a target is small, the neutron yield at high energy will be more.
So if one wants to get high energy neutrons, low-Z material is a good choice for the target material.

\begin{figure}[htb]
	\centering
	\includegraphics[width=8.6cm]{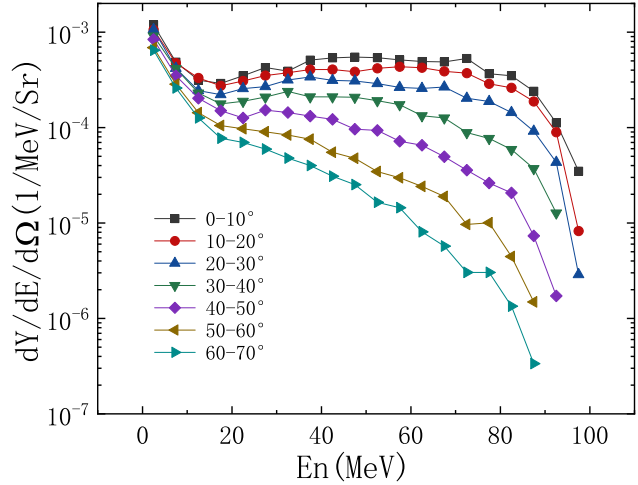}
	\caption{Comparison of neutron yield at different angles. The deuterium beam at 100 MeV reacts with a carbon target. The target thickness is 4 cm, and the target radius is 7 cm. }
	\label{fig-6}
\end{figure}

Figure \ref{fig-6} shows the angle distribution of neutrons for carbon target.
The target thickness is 4 cm, and the target radius is 7 cm.
The neutron spectrum is recorded every 10$^{\circ}$.
It can be found that as the angle increases, the high energy neutron yield decreases.
Between 0-10$^{\circ}$, the number of high energy neutrons is highest.
On the contrary, it can be seen that between 60-70$^{\circ}$, the high energy neutron yield is the smallest.
In the low energy neutron region (0-10 MeV), it can be found in Fig. \ref{fig-6} that as the angle increases, the number of low energy neutrons decreases.
For d + $^{12}$C reaction, the neutron yield is greater within a small angle range.
The result is consistent with the cross sections in Fig. \ref{fig-2}.

\begin{figure}[htb]
	\centering
	\includegraphics[width=8.6cm]{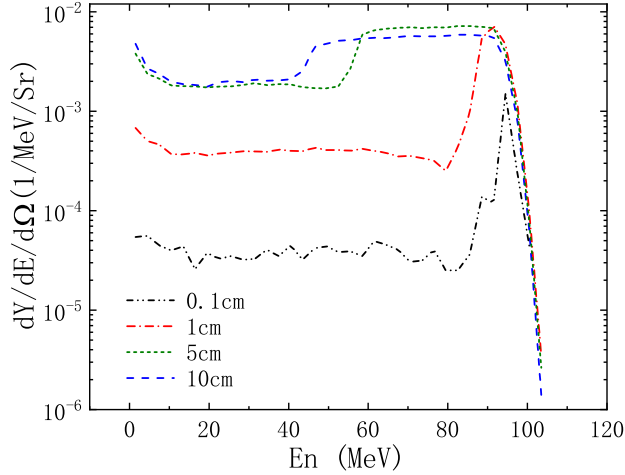}
	\caption{Neutron energy spectrum of different target thickness for reaction d+$^{12}$C at 200 MeV. }
	\label{fig-7}
\end{figure}

Figure \ref{fig-7} shows the neutron energy spectrum for different target thickness.
The detection range of the detector is set at 0 -10$^{\circ}$.
Carbon is used as the target material.
It can be found that as the thickness of the target increases, the yield of neutrons will increase.
In Fig. \ref{fig-7}, these curves all reach their peaks at the neutron energy of around 95 MeV, which is near half of the incident deuteron energy.
The deuteron is composed of a neutron and a proton.
The neutron that make up the deuteron has half of its energy.
During the spallation reaction of d + $^{12}$C, the deuterium nucleus will break up and release neutrons.
So most of the neutron energy detected by the detector is around 100 MeV.
The peaks of the curves corresponding to the thickness of 5 cm and 10 cm are not obvious.
For thick targets, the neutrons generated by the spallation reaction will be slow down in the target.
The deuteron after slow down also produce the neutrons, the energy of which is smaller than 100 MeV.
Thus, if one wants to obtain quasi-monoenergetic neutron, the thin target is a good choice.

\begin{figure}[htb]
	\centering
	\includegraphics[width=8.6cm]{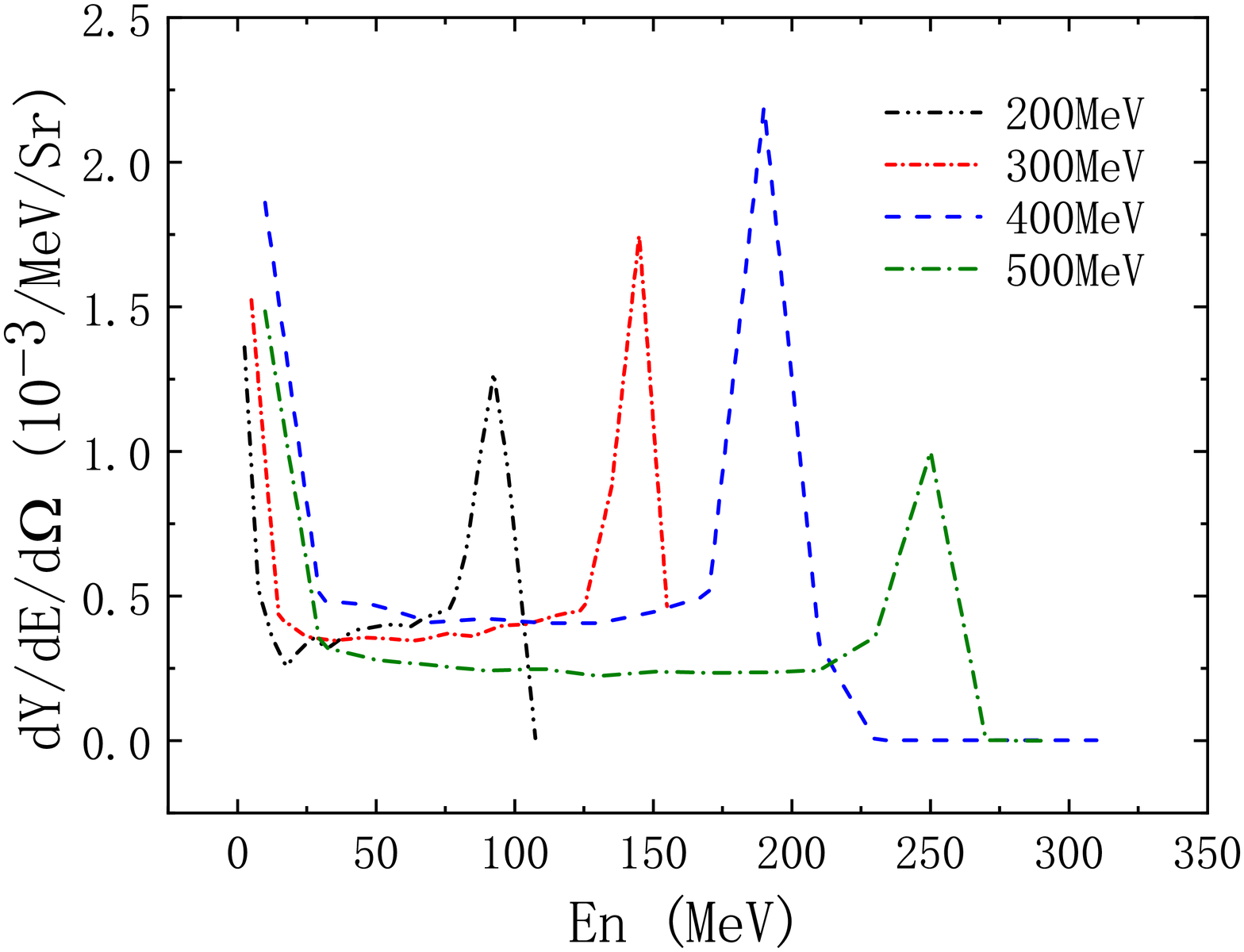}
	\caption{Neutron spectrum at 0-17$^{\circ}$ from the interaction between deuteron beam and carbon target with thickness 0.1 cm. The energy of the incident deuteron is 200, 300, 400, and 500 MeV.}
	\label{fig-8}
\end{figure}

Figure \ref{fig-8} shows the neutron spectrum caused by deuteron with different energies.
One sees that the quasi-monoenergetic spectra are obtained.
As the energy of the incident deuteron increases, the neutron energy corresponding to the peak also increases.
It can be noticed that the neutron energy corresponding to the peak is half of the energy of the incident deuteron, which is consistent with the result of Fig. \ref{fig-7}.
So if we want to get high energy neutrons, we should choose a deuterium beam with an energy that is twice the energy of the neutrons.

\subsection{Comparison of neutron yield by Geant4 and analytical method}

\begin{figure}[htb]
	\centering
	\includegraphics[width=12cm]{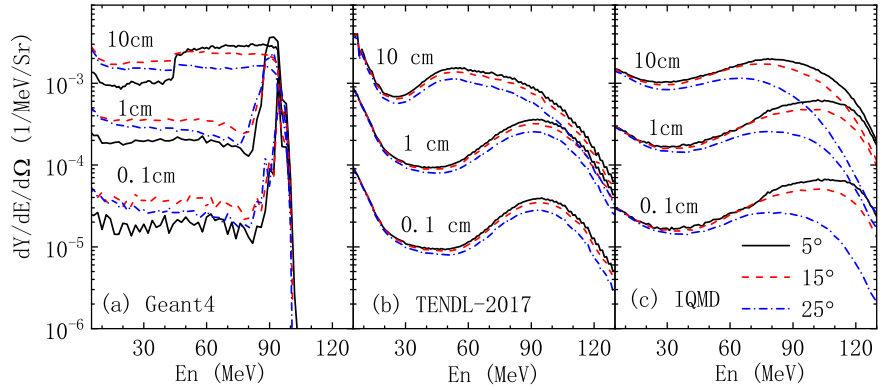}
	\caption{Comparison of neutron energy spectrum from interaction between deuteron beam at 200 MeV and carbon target calculated by Geant4, and analytical methods within TENDL-2017 and IQMD cross sections. }
	\label{fig-9}
\end{figure}

In the above subsection, the calculation by the Geant4 has shown that interactions between deuteron beam and thin carbon target can produce the quasi-monoenergetic neutrons.
However, there is considerable uncertainty of the neutron productions cross sections in the high energy region.
In order to study the propagation of the uncertainty, the analytical calculations are performed within TENDL-2017 and IQMD cross sections.
Figures \ref{fig-9}(b) and \ref{fig-9}(c) show the neutron yields from the interaction between deuteron beam at and carbon target calculated within TENDL-2017 and IQMD cross sections.
They are compared with the calculations by Geant4 in Fig. \ref{fig-9}(a).
In the calculation, the target thickness 0.1, 1, and 10 cm are applied respectively.
The emission angles 5$^{\circ}$, 15$^{\circ}$, and 25$^{\circ}$ are considered.
The thickness and angle dependence are similar for the calculations by the three methods.
The neutron yield increases with increasing thickness, but decreases with increasing angle.
However, differences are obvious for results by the three methods.
Calculations by the Geant4 show narrow peaks for target thickness 0.1 and 1 cm, which indicates the possibility to produce the quasi-monoenergetic neutrons.
Peaks can been observed for the calculations by the analytical methods, but are wider than those by Geant4.
The angle dependences of the calculations by the two analytical methods are also different.
The analytical calculations within the TENDL-2017 cross sections show weak angle dependence, which indicates that one can use the high energy neutrons at the wide emission angle.
The strong angle dependence by the IQMD cross sections enhances the difficulty to use the neutrons.
Because the small angle, where the neutron yield is huge, has been occupied by the incident deuteron beam.

\begin{figure}[htb]
	\centering
	\includegraphics[width=8.6cm]{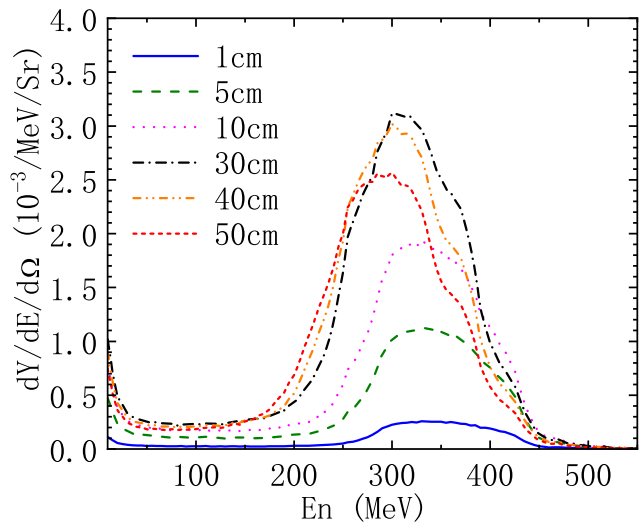}
	\caption{Neutron yields at 0-10$^{\circ}$ from interaction between deuteron beam at 800 MeV and carbon target with different thickness. }
	\label{fig-10}
\end{figure}
In the energy region higher than 200 MeV, the TENDL-2017 cross sections are unavailable.
Thus, we will only show the neutron yield by the analytical method by the IQMD cross sections for the deuteron energy higher than 200 MeV.
Figure \ref{fig-10} shows the neutron yields at 0-10$^{\circ}$ from interaction deuteron beam at 800 MeV and carbon target with different thickness.
One sees the the obvious peaks for high energy neutron yields.
With increasing thickness, the peak position decreases but the peak value rises and then fails.
The peak is the highest when the thickness is 30 cm.
When the target thickness is larger than 30 cm, the peak value begins to decrease.
The peak values corresponding to 40 cm and 50 cm are smaller than that of 30 cm.
The carbon target with thickness 30 cm is the best for produce high energy neutron for deuteron beam at 800 MeV.

\section{\label{con} Conclusion}

In summary, by comparing with the experimental data and the predictions in other databases, the  isospin dependent quantum molecular dynamics (IQMD) model is proved to be reliable to predict the neutron production cross sections in proton-induced reaction at 256 MeV.
Then the IQMD model is applied to calculate the high energy neutron cross sections in deuteron-induced reactions, the experimental data of which are still very scarce.
By comparing the calculations with the predictions by TENDL-2017, ENDF/B-VI.8-HE, and JENDL/HE-2007 databases, it is indicated that the uncertainty of the cross sections needed for the simulation of the deuteron beam interaction with matter is still huge.

The Geant4 (version G4.10.05p.01) is applied to simulate the deuteron beam interaction with matter.
Meanwhile, the analytical method is developed to calculate the yield of the forward neutrons emitted in the interaction between the deuteron beam and thick target.
The cross sections taken from the TENDL-2017 database and calculated by the IQMD model are used as input.
The results of analytical method are compared with those calculated by the Geant4.
The differences appear for the peak positions and peak width, since the uncertainty of the cross sections is still huge.
However, the similar trend indicates that the neutron yield at half energy of the deuteron beam is considerable.
If one wants to get high energy neutrons, low-Z target and small emission angle may be considered.
The neutron yield at half energy of the deuteron beam increases with the increasing thickness of the target.
But the thin target is a good choice if one wants to obtain quasi-monoenergetic neutron.

\section*{Acknowledgments}

This work was supported by the National Natural Science Foundation of China under Grants No. 11875328.

\bibliography{mybibfile}

\end{document}